# A Step towards Advanced Metering for the Smart Grid: A Survey of Energy Monitors

Anwar Ul Haq, and Hans-Arno Jacobsen

*Abstract*--The smart grid initiative has encouraged utility companies worldwide to rollout new and smarter versions of energy meters. Before an extensive rollout, which is both labor-intensive and incurs high capital costs, consumers need to be incentivized to reap the long-term benefits of smart grid. Off-the-shelf energy monitors can provide consumers with an insight of such potential benefits. Since energy monitors are owned by the consumer, the consumer has greater control over data which significantly reduces privacy and data confidentiality concerns. We evaluate several existing energy monitors using an online technical survey and online product literature. For consumers, the use of different off-the-shelf energy monitors can help demonstrate the potential gains of smart grid. Our survey indicates a trend towards incorporation of state-of-the-art capabilities, like appliance level monitoring through load disaggregation in energy monitors, which can encourage effective consumer participation. Multiple sensor types and ratings allow some monitors to operate in various configurations and environments.

*Index Terms*--Demand side management, disaggregation, energy monitor, non-intrusive load monitoring, smart grid.

## I. INTRODUCTION

THE recent transition in energy policy from traditional fossil fuels to cleaner renewable energy resources has fostered the wide acceptability of green energy technology [1]. Smart grid is expected to embed digital intelligence in our electrical system from generation to transmission, distribution, consumption, and pricing of electrical energy. The discussions regarding smart grid are often centered on energy generation where most of energy saving is expected [2]. Apart from generation, considerable savings can be achieved through proper monitoring and planning of the efficient energy usage through demand side management (DSM) [3], [4].

The integration of weather-driven renewable energy into the system requires more flexibility in generation, distribution, and utilization [5]. To some extent, this flexibility can be achieved through the advanced metering infrastructure (AMI) providing better load forecasting and anomaly detection [6]. Consumers play an important role as renewables are expected to increase on the demand side. Being at the heart of any smart grid infrastructure, consumers need to gain adequate understanding of smart metering solutions. Consumers can be incentivized to participate effectively if provided with adequate information to help them understand the potential advantages of smart grid.

The primary aim of any household or building manager is to intelligently utilize appliances with regard to user comfort and preferences while emphasizing energy efficiency. To manage the amount of energy spent, it is necessary to measure how and where this energy is consumed. Under the smart grid paradigm, the next generation smart buildings will require bi-directional power and data communication to reduce demand during high wholesale market price or grid malfunction [7]. This situation calls for highly interactive metering technologies that act as middleware to seamlessly gather data regardless of vendor or communication protocol.

The concept of electricity monitoring emerged immediately after the inception of electricity generation and distribution systems during the late 18th century. The first commercial use of electricity was direct current (DC) and electro-chemical meters were introduced initially to measure electricity consumption [8]. These meters were labor-intensive as they required periodic removal and weighing of plates from an electrolytic cell. Electro-chemical meters were then replaced by electro-mechanical meters, also known as induction meters or Ferraris meters [9]. The early electro-mechanical meters measured charge in ampere-hours and calculated energy consumed during the billing period.

In the beginning, electricity was primarily utilized by lighting systems and to a lesser extent for operating electric loads like electric motors. As more industries shifted from oil and gas to electricity, there was an enormous increase in energy demand and hence the need to measure electricity use accurately. Modern buildings, both residential and commercial, constitute a major portion of electricity demand. It is estimated that around 73% of electricity in the US is consumed by buildings [10]. During 1999 to 2004, the consumption of electricity in the EU alone has increased by 10.8% [11]. In Europe, the energy consumption in buildings accounts for only 15% of electricity use with the major portion of energy supplied by fossil fuels, making these buildings environmentally irresponsible [12].

With this paper, we explore the different energy monitors (e-monitors) currently available to consumers. The main goal of our work is to help researchers, building managers and consumers choose the monitor best suited for their specific applications. Although some of these e-monitors are appropriate to manage renewables and can provide added features, such as load disaggregation for appliance level monitoring, they are often overlooked due to a lack of

Anwar Ul Haq and Hans-Arno Jacobsen are with the Technical University Munich, Department of Computer Science, 85748 Garching, Germany (e-mail: anwar.haq@tum.de, jacobsen@in.tum.de).

2available technical data about their capabilities. Unlike smart meter, which is owned by the utility, the e-monitor is bought and managed by the consumer. The e-monitor allows consumers to have more control over data and in the future, the consumer can share non-private data collected by the e-monitor with the utility to facilitate in load-forecasting.

The rest of the paper is organized as follows. Section II describes the background of our work with relevant technical details while Section III compares different e-monitors and Section IV includes key findings and suitable suggestions for choosing e-monitors based on application. In Section V we draw conclusions on the comparisons of the monitors.

## II. BACKGROUND

Before going in detail, it is necessary to differentiate between a smart meter and an e-monitor. A smart meter is the next generation meter capable of linking a house or building with a utility company by enabling two-way communication and power exchange between them [13]. A smart meter also assists in remote billing and instant load feedback to the utility for load forecasting. As it is owned by the utility, the smart meter comes with inherent drawbacks related to data confidentiality and privacy [14]-[16]. On the contrary, the e-monitor is owned by the consumer and can work with existing meters without any direct effect on billing. The e-monitors have user-friendly interface and can be easily installed by clipping them around a current carrying wire or directly inserting them in a power plug. Due to local and private cloud storage, the e-monitor can minimize the privacy concerns and added features, such as disaggregation and efficient integration of renewables, can encourage consumers and building managers to effectively participate in DSM.

For a fair comparison it is important to view how different vendors and platforms measure and calculate energy consumption. For load monitoring, there are two main categories of e-monitors available on the market; *single point* and *multi-point* e-monitors. The *single point* e-monitors capture the aggregate energy consumption of the whole house, building or industrial facility. The *non-invasive* or *non-intrusive* e-monitors are clamped around the current carrying wire which makes them easy and safe for customers to install themselves. The *multi-point* e-monitors capture constant data at several locations and are preferred for detailed load monitoring, even down to monitoring power usage of individual appliances. Monitoring of appliances can result in more engaged consumer participation as they are able to better identify power hungry appliances and limit their use during peak hours.

Recently, the trend towards the use of *non-intrusive load monitoring* (NILM) has gradually increased [17]-[19]. NILM refers to the single point sensing of the electricity mains to gather overall energy consumption. By using a high-resolution data set, the cumulative load is disaggregated to identify each constituent appliance and estimate its power consumption [17]. This approach takes advantage of advances in computer science and provides an intelligent method of obtaining appliance specific load profile instead of using a dedicated device to monitor each individual appliance, also known as intrusive load monitoring (ILM). NILM monitors the aggregate load and scans for load signature, a unique feature for each appliance [20].

For a fair comparison of e-monitors, we have outlined six dimensions including types of parameters, sampling frequency, accuracy, resolution, application area, and cost of monitoring equipment on which we base our comparisons.

### A. Parameter Type

Except voltage and current, most of the main parameters (if utilized) are derived and calculated using standard mathematical formulations. These parameters are derived internally and for the most part, a subset of these parameters is utilized and displayed to consumers.

#### 1) Voltage Waveform (V):

The potential to drive electrons between two points is called voltage and is one of the basic parameters to examine the system stability. The voltage waveform is measured instantaneously and is utilized for calculation of further parameters. Voltage waveform measurement assists to make corrective measures against harmful low and high voltage levels. Usually the voltage transformers (AC-AC adaptors) are used to measure the peak and root-mean-square (RMS) voltage of the line.

#### 2) Current Waveform (I):

To measure the flow of electrons across a conductor, current measuring sensors (e.g., current transformer, hall effect transducers, Rogowski coils, etc.) are deployed. Sometimes shunt (consisting of a material with accurately known resistance to calculate voltage corresponding to passing current) and optical measurement for pulse counting (e.g., LED flashes) are also used to measure current. Unlike voltage waveform, the current waveforms are not stable sine waves; they vary considerably depending upon the type of load attached as illustrated in Fig. 1. Each load type (resistive, inductive or capacitive load) has a different influence on the current curve and often the inrush current features are used for appliance segregation using NILM. The current waveform can be an important factor to determine system load type where more resistive loads tend to have a stable effect on curve.

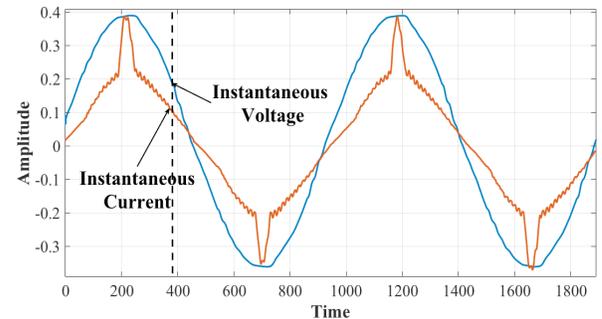

Fig. 1. Instantaneous voltage and current waveforms

#### 3) RMS Current:

The process of measuring alternating current (AC) is more complex as compared to direct current (DC). This is due to the fact that AC continuously alternates between positive and negative peak as a sine wave and its average is zero. So the most common way to measure current is to express its effective value, also known as root mean square (RMS). RMS is obtained by dividing the peak current ($I_P$) with the square root of 2. Mathematically,

$$I_{RMS} = \frac{I_P}{\sqrt{2}} \quad (1)$$

The above formula assumes a pure sine wave which is not exactly the case in reality, especially for current. A realistic picture of the voltage and current measurement at an instant is indicated in Fig. 1. A better way of handling real-time current is by taking the square root of the mean squared instantaneous current values, varying continuously and averaged over the complete cycle. Mathematically,

$$I_{RMS} = \sqrt{\frac{\sum_{n=0}^{N-1} I^2(n)}{N}} \quad (2)$$

Where $I(n)$ = Current at instant n
$N$ = Number of samples per cycle

*4) RMS Voltage:*

The effective or RMS voltage is more stable as compared to RMS current and can be calculated similarly using the formula

$$V_{RMS} = \sqrt{\frac{\sum_{n=0}^{N-1} V^2(n)}{N}} \quad (3)$$

Where $V(n)$ = current at instant n
$N$ = number of samples per cycle

*5) Real Power:*

The main feature used by almost every energy-metering device is the real power. It is the actual rate at which energy is used and is calculated through the voltage and current measurements. Mathematically,

$$P = VI \text{ or } P = I^2 R \text{ or } P = \frac{V^2}{R} \quad (4)$$

The real power is usually defined as the power consumed by the resistive load.

*6) Power Factor:*

Power factor is used to distinguish between resistive, inductive and capacitive appliances. It determines the phase difference caused by the inductive and capacitive components. A positive phase angle indicates net inductive reactance of the circuit where the current lags voltage. On the contrary, a negative phase angle indicates net capacitive reactance of the circuit as the current leads voltage. Mathematically, it can be expressed as

$$Power\ factor = \frac{\operatorname{Re}al\ power\ (P)}{Apparent\ power\ (S)} \quad (5)$$

Here, the apparent power is the vector sum of real and reactive power (power associated with inductance and capacitance). Mathematically, it can be calculated as

$$Apparent\ Power\ (S) = V_{RMS} \times I_{RMS} \quad (6)$$

The power factor determines the need to deploy the correction capacitors and reduce energy waste [21].

*7) Harmonics:*

Harmonics are considered as signals whose frequency is the whole number multiple of some fixed frequency, commonly known as the fundamental frequency or the first harmonic. These harmonics or higher frequency components occur due to pulsating devices (such as frequency drive, electric welders, etc.) resulting in system heating and overvoltage [21]. For instance, the third harmonic of a power supply operating at 50 Hz will be 150 Hz. The harmonics are created by different electronic components present in the appliance circuitry. This produces a new distinct waveform due to superposition of different harmonics.

Fast Fourier transform (FFT) resolves the superimposed waves into their constituent waves. In energy monitors, another term commonly associated with harmonics is the total harmonic distortion (THD). It refers to the presence of harmonic distortion caused by the non-linear loads. THD determines the power quality of the system where a lower THD indicates a reduction in heating, peak currents and losses [22]. Due to these distinctive features, the higher order harmonics are useful for power disaggregation application.

*8) Wavelet Transform:*

Wavelets are essentially short bursts of waves that die quickly. The Fourier transform is mostly utilized as a feature to describe a wave based on its superior encoding, filtering, compression and noise reduction capabilities. The only limitation is caused by the Heisenberg uncertainty principle. This uncertainty affects the resolution of the signal: the signal can be accurate in time domain or frequency domain but not in both. Wavelet transform overcomes uncertainty by adding the signal into wavelets and limiting the wavelets in time and frequency [23]. In turn, because the wavelets are comprised of finite waves, they provide more resolution in time. In this way, frequency is also handled in time by changing the scale of the wavelet.

*9) Load Profile:*

The load or energy consumption profiles help determine the pattern of energy usage with respect to time. For consumers, these patterns are helpful in finding energy leakage. Utility companies use these patterns as a statistical tool for load forecasting [24]. Very precise and appliance level consumption profiles can be produced either through distributed approach, using smart plugs, or through load disaggregation techniques using NILM [25]. In addition to calculating power consumed by the appliance at a particular point in time, the shape of the power consumption profile also provides useful information. For instance, during the operation of an appliance, the number of peaks can be uniquely identified and associated with a particular appliance or for grouping similar appliances. Consumption profiles are useful in distinguishing the multistate appliances based on identical patterns of peaks during their operation.

*10) Eigenvalues (EIG):*

The current waveform for dynamic loads, such as air-conditioners, varies from cycle to cycle. In order to capture these variations, we usually perform eigenvalue analysis by rearranging time series current waveform into matrix form. The study on appliance load signature [26] indicates that power hungry appliances usually have higher 1$^{st}$ EIG features.

Even the 2nd and 3rd EIG features of these appliances show good correlation and can be utilized as a feature for appliance identification.

*11) Instantaneous Admittance Waveform:*

Instantaneous Admittance Waveform (IAW) is defined as the quotient between the current and voltage waveform at a specific point in time. Mathematically [27],

$$IAW(t) = \frac{I(t)}{V(t)} \quad (7)$$

As most household appliances are connected in parallel with respect to the main circuit, it is more common to choose admittance instead of impedance at any point in time. This simplifies the calculations because small differences in impedance are harder to observe as compared to admittance (inverse of impedance) [26]. Although IAW proved to be a robust feature [28], it can introduce some numerical instability due to sharp spikes as the voltage approaches zero.

*12) Side Channel Features:*

In addition to the main features discussed above, external parameters are also helpful in energy monitoring of individual appliances and are known as side channel features. Features like time of day, weather information, appliance location in the circuit (single or three phase), and appliance usage pattern can help boost the appliance detection process [29]. The side-channel assisted NILM can greatly enhance the ground truth verification capabilities of the NILM-based systems. In addition to external parameters, light, sound and electro-magnetic field (EMF) are utilized as side channel features. To obtain the appliance switching information, the electro-magnetic sensors are placed in close proximity of the appliance under observation (usually 5-10 cm range) [30]. Similarly, channel electrical noise can also assist in appliance identification but has a strong dependency on the electrical wiring system, the main drawback of using this approach [31].

*B. Resolution*

Since voltage and current signals are continuous in nature (analog), to calculate the other set of features (e.g., real power, RMS voltage and current, power factor, etc.), analog signals need to be converted to digital signals. The uncertainty in the digital signal is determined by the closeness of measurement points in the analog input. This uncertainty is known as the resolution of the signal. Hence, sampling frequency and resolution are typically interdependent. Whereas the resolution on the x-axis (time) is set by the sampling frequency, on the y-axis it is determined by the number of bits in the variables used to represent each value (bits of ADC). The resolution or smallest step size of n bit ADC with $2^n$ distinct step levels can be calculated through least significant bit (LSB) as

$$1\, LSB = \frac{FSR}{(2^n - 1)} \quad (8)$$

Where $LSB$ = Least significant bit
$FSB$ = Full scale reading
$n$ = number of bits of ADC

*C. Sampling Frequency*

For an oscillating signal, frequency refers to the number of complete cycles per second. For metering, the choice of any specific sampling frequency or sampling rate of voltage and current signals depends upon the amount of information we are interested in obtaining from these signals. The sampling frequency may range from hourly reading to very high frequency (MHz) range.

To observe the harmonics and transient switching response of the appliances, it is better to utilize higher sampling frequency. According to Nyquist criterion, one must sample twice as much as the highest harmonic required for the measurement [18]. For any meter, the minimum sampling rate is 16 samples per cycle to calculate the RMS values. This rate increases to a minimum of 64 for calculating total harmonic distortion (THD) and 256 samples per cycle for harmonic spectrum analysis [21]. The sampling done for analog to digital conversion might not be the same as samples reported for display. Most of the modern e-monitors down sample to lower sampling rates in order to reduce storage requirements.

As a general rule for disaggregation, higher sampling frequency allows us to distinguish more appliances in near real-time. For example, with a sampling frequency of 10 to 40 kHz, one can differentiate 20 to 40 appliances. Increasing the sampling frequency beyond 1 MHz can help distinguish 40 to 100 unique appliances [32].

*D. Accuracy*

Accuracy is determined by closeness of measured values with the actual value. A study on commercial smart meters indicates accuracy around 99.96% within a +/- 2% accuracy range [33]. Generally, accuracy is considered as the most specified feature for any meter and often 0.5% minimum accuracy is considered adequate for revenue billing [21].

The inaccuracies mainly stem from the analog to digital converters (ADC) and transformers (both voltage and current). The ADC's introduce a quantization error, which can be reduced using a higher bit ADC corresponding to the smaller step size. Usually, 1 least significant bit (LSB) is used as a reference, which defines the width of each step. On the contrary, the current transformer (CT) shows best performance when it is operating near its full load.

*E. Application*

The e-monitors are utilized almost everywhere electricity monitoring is required – how they are utilized differs based on the area of application. The residential sector consists of housing units, the commercial sector consists of non-manufacturing business establishments (e.g., warehouses, hotels, restaurants, etc.,) and the industrial sector consists of manufacturing units with fixed machinery (e.g., motors, drives, generators, etc.) [34]. In residential and commercial buildings, the aggregate load is mostly monitored using electro-mechanical meters. This single point sensing can be single- or three-phase monitoring at the whole house or building level. If one is interested in more detailed energy consumption information, circuit level monitoring can be applied which can be termed as multipoint energy sensing.

Recently, interest is growing in appliance level energy monitoring to help consumers view fine-grained energy consumption information at appliance level. The NILM approach utilizes single point sensing and using some machine learning algorithms helps disaggregate energy data to identify appliance-specific energy consumption [17]-[19].

*F. Cost*

Cost is one of the most important factors when purchasing an e-monitor. A large-scale longitudinal survey was carried out by the Department of Energy and Climate in the UK to estimate the cost of different monitoring solutions for electricity and gas. The survey results recommend three different energy monitoring packages ranging from £ 210 to £ 950 per dwelling [35].

### III. Survey and Comparison Results

For this research, a comprehensive online survey [36] of various energy monitoring solutions available on the market was conducted. The purpose of conducting the online survey was to obtain detailed technical information regarding e-monitors as limited information is publically available. A total of 54 different companies were shortlisted and invited to participate in the survey. The survey included 79 different e-monitors, all of which are off-the-shelf monitors and hence owned directly by the customers. We received responses from 18 companies for 27 e-monitors through the online forms, a response ratio of 34.1%. We further collected information from 9 companies on their 14 e-monitors through online literature. For three respondents, we were not allowed to publish the data but their information is included in the results. In the survey, we grouped similar monitors from the same vendor together. For complete data, please refer to our technical note [37].

*A. Application Area*

We identified the application of the available e-monitors and divided them into three main categories; residential, commercial (including buildings, offices) and industry. More than 90% of the e-monitors can be utilized in the residential sector and more than 60% can be utilized for commercial use. Similarly, more than 30% of the e-monitors can be utilized for industrial use. Besides different load types, some e-monitors can be deployed in multiple environments.

*B. Device Type*

For our survey, we categorized different monitors based on their installation and measurement position in the electrical network of buildings. These include smart plugs, which are mounted on the wall outlet to measure individual end appliance. For testing load disaggregation algorithms, these plugs are utilized for the collection and validation of turn on/off events to establish ground truth. The smart energy meters installed between the electricity mains and distribution box inside the building, like smart-me, are owned and controlled directly by the customer and are included in the survey. The majority of e-monitors included in this study are installed in the fuse box or attached directly to the electric main and meters. These e-monitors usually incorporate electricity monitoring and analytics aimed at reducing the monthly electric bill through effective customer participation. This study also includes the gateways installed between the e-monitoring unit and the internet to upload information directly to a cloud. This survey indicates that e-monitors makeup over 75% of the market followed by the smart plugs (Fig. 2).

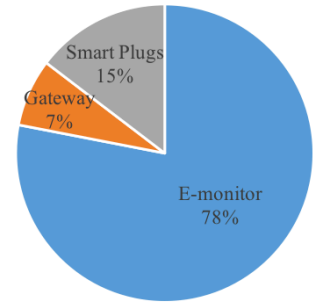

Fig. 2. Different categories of e-monitors

*C. System Type*

We surveyed the different monitoring solutions and their compatibility with either single- or three-phase systems. According to our survey, over 70% of e-monitors are compatible with both single- and three-phase systems (Fig. 3). The single-phase e-monitors can be scaled to three phases by using multiple units and can be calibrated using pure resistive load so that the voltage and current curves don't mismatch.

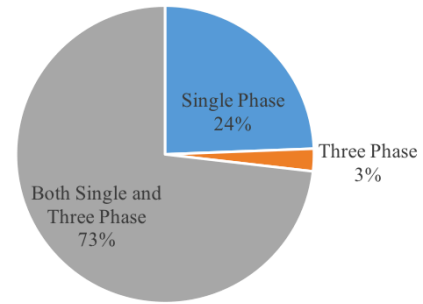

Fig. 3. E-monitor utilization system

*D. Sensor Type*

Some e-monitors only measure current of the system while assuming the constant voltage; for three-phase systems, it is essential to include the voltage for at least one phase if not for all phases. The three phases are ideally considered balanced but if the load is not evenly distributed in each phase, which is very often the case, then different phases tend to have different voltages. The survey results indicate that nearly 60% of the monitoring systems use current transformers (CT) and some utilize Rogowski coils for the measurement of current (Fig. 4).

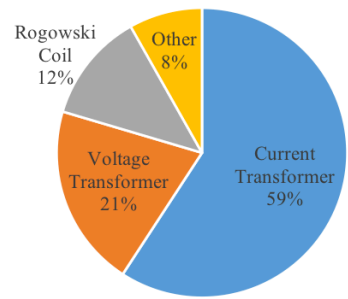

Fig. 4. Types of sensors used by e-monitors

Although the Rogowski coils offer a broad range of measurement windows and are safer to use than normal CTs, they are still underutilized. Higher cost, as compared to the current transformer, is an important factor for the under deployment of Rogowski coils. On the other hand, only about 20% of the e-monitoring solutions independently measure voltage. Some monitoring solutions directly use shunt, pulse count, and optical measurement from the meter.

*E. Sensor Rating*

The type of sensor utilized depends on the application and is defined by the maximum load to be measured. Current transformer consists of an iron core with primary and secondary coils wrapped around it. The survey results indicate a wide variety of current transformers utilized by different e-monitors (Fig. 5). The results also indicate that around 70% of current transformers are rated up to 200 A. This is due to extensive use of e-monitors in the residential sector where the maximum load at any given time does not exceed 200 A.

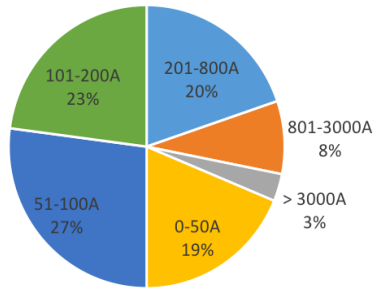

Fig. 5. Ratings of current transformers

*F. Parameters*

The e-monitors primarily measure the system voltage and current passing through a point at any given point in time. Based on these measurements, many different parameters can be calculated. According to survey results, single parameters are used most commonly followed by the use of 5 or more parameters (Fig. 6). Although almost all of the e-monitors measure the voltage and current (except when voltage is assumed constant), these measurements are not necessarily displayed to the user.

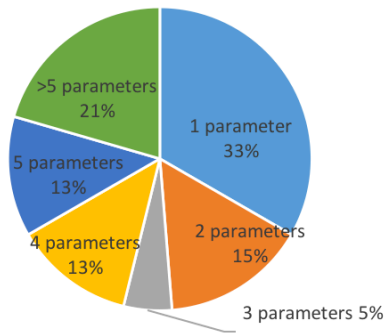

Fig. 6. Number of parameters used by e-monitors

Approximately 80% of e-monitors utilize current and real power to indicate load, followed by voltage (Fig. 7). When considering load disaggregation, it is always better to incorporate more parameters as certain parameters work better for particular load types [17], [38]. The inclusion of load specific parameters enhances the distinction among the appliances. In our survey, some e-monitors utilized up to 9 distinct parameters.

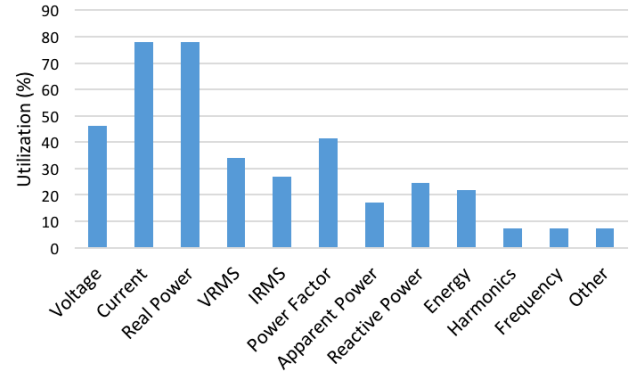

Fig. 7. Parameter types used by the e-monitors

*G. Sampling Frequency*

Sampling frequency is an important dimension for comparison and is required for proper conversion of analog signal to digital. The voltage waveform is usually quite stable and can be reconstructed easily, but the current waveform is not even close to a proper sine-wave and hence requires increased sampling for proper digital reconstruction.

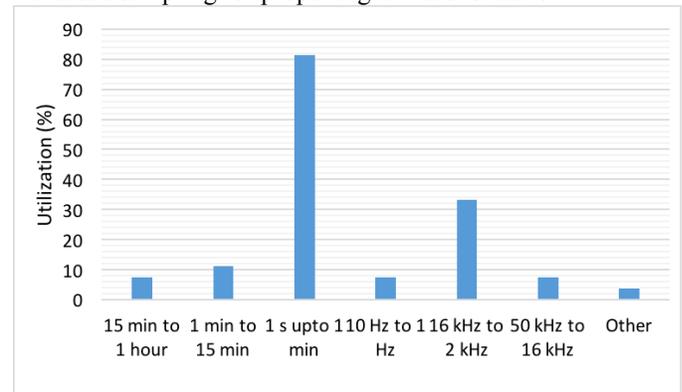

Fig. 8. Sampling frequency used by e-monitors

All e-monitors have sufficient sampling rate to satisfy the Nyquist criteria but raw samples are down-sampled and made available to the consumers. Most commonly, a 1 s to 1 min sampling rate is made available by e-monitors (Fig. 8). Armel et al. [32] suggest that high sampling frequency can lead to real-time appliance identification.

*H. Resolution (power and bits)*

The resolution is determined by the number of bits of analog to digital converter (ADC) and defines the number of codes that can be formed digitally using these bits. Not all the participants disclosed the number of bits but most ADC lie between 10 to 16 bits. From the received data, most of the monitors were using 16-bit ADC. Another important parameter associated with resolution is the power resolution, i.e., minimum level of power measured by these appliances. Most e-monitors have a power resolution between 1 W to 5 W making them capable of very precise and accurate metering.

*I. Measurement Channels*

Some load appliances rely on three-phase measurement while others operate on single phase consequently affecting the number of channels that can be monitored.

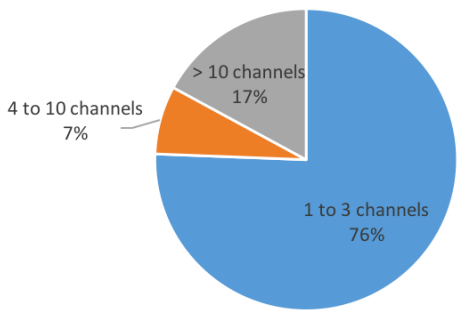

Fig. 9. Number of channels measured by e-monitor

Greater than 75% of appliances support the measurement of one to three distinct channels (Fig. 9). With three input channels, one can either measure three separate single phases or one three-phase. With *Verdigris*, one can accurately measure about 42 different channels/circuits [39]. *GridSpy* is another example of a system capable of measuring 6 circuits per *node* (wireless data collector), 30 circuits per *hub* (collects and uploads data) and can scale up to 600 circuits per *site* [40]. *CURB Pro* is also capable of monitoring 18 breakers per hub and this breaker level measurement (hardware disaggregation) facilitates disaggregation algorithms as the type of load appliance on a particular breaker is already known [41].

*J. Storage*

The e-monitors are capable of storing data either locally or uploading to a cloud to perform further analytics. Most e-monitors prefer to upload data to a cloud while some have dual capability to store data locally and, at the same time, upload it to a cloud (Fig. 10). Issues such as data privacy and confidentiality can be decreased by using local or private cloud storage. With such data, arrangements can be made for the consumer to take advantage of load disaggregation and compute appliance level power consumption.

*K. Cost*

To compare costs, we converted all prices into Euro. According to our survey, the cost of e-monitors varies between €38 and €3,220 for a single product according to its application. The average price range is €435 to around €600, depending upon single or three phase systems and accessories utilized with the e-monitors. The prices of smart plugs range between €15 and €180 with an average price of €77.

## IV. KEY FINDINGS

The main purpose of this technical survey is to provide information to facilitate researchers, facility managers, and general consumers in selecting a monitoring system that best fits their requirements. Commonly, e-monitors are utilized to track and display the energy used and saved. The key differences among e-monitors include application area, sampling, resolution, system configuration and sensor type.

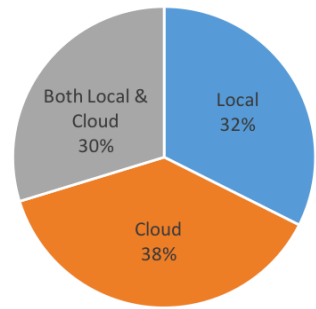

Fig. 10. Different storage options for e-monitors

We believe consumers can participate efficiently for demand response (DR) once they are provided with real-time energy consumption information, especially at appliance level. Information on energy consumption can help identify energy hungry appliances and eventually facilitate in predictive maintenance. NILM is a technique that relies on single point sensing and uses machine learning techniques to estimate the power consumption of each appliance. While some of the surveyed e-monitors, such as Smappee, Neurio, Verdigris, and Curb, already utilize these disaggregation techniques, most of the other e-monitors possess enough resolution, parameter diversity, and sampling frequency to use disaggregation.

Similarly, in addition to being a labor-intensive task, one of the biggest hurdles in the speedy rollout of smart meters are data confidentiality issues for consumers. One way to demonstrate the advantages of smart grid is by letting the consumers experience such benefits. Due to private storage and ownership of both e-monitor and data, consumers can essentially experience smart grid benefits. By using disaggregation algorithms with this data, one can observe appliance level power consumption. Most e-monitors can also be used in multiple settings and configurations as they come with numerous options in terms of sensor rating and application area.

## V. CONCLUSIONS

The goal of this study was to compare different state-of-the-art e-monitors available on the market. Through online technical survey and detailed product review, we compared 41 energy monitors based on several dimensions including measured and derived parameters, sampling frequency, accuracy, resolution, area of application and cost. The comparison suggests that most e-monitors have a tendency to incorporate advanced features and upgrade existing meters into next-generation intelligent metering units. In the future, these intelligent meters can act as the point of contact between smart buildings by grouping them for local demand response. Before complete rollout of smart meters, consumers can realize their considerable advantages through selecting and using intelligent off-the-shelf e-monitors.